\documentclass[preprint,prl,showpacs,superscriptaddress]{revtex4-1}
\usepackage{amsmath}
\usepackage{graphicx}
\usepackage{amssymb}
\usepackage{times}
\usepackage{color}
\usepackage{multirow}
\definecolor{lr}{rgb}{1.0,0.3,0.3}
\definecolor{dg}{rgb}{0.0,0.5,0.0}

\usepackage{comment}

\graphicspath{./}
 


\begin{document}

\title{Identification of Si-vacancy related room temperature qubits in 4H silicon carbide}

\author{Viktor Iv\'ady}
\email{vikiv@ifm.liu.se}
\affiliation{Department of Physics, Chemistry and Biology, Link\"oping
  University, SE-581 83 Link\"oping, Sweden}
\affiliation{Wigner Research Centre for Physics, Hungarian Academy of Sciences,
  PO Box 49, H-1525, Budapest, Hungary}

\author{Joel Davidsson} 
\affiliation{Department of Physics, Chemistry and Biology, Link\"oping
  University, SE-581 83 Link\"oping, Sweden}

\author{Nguyen Tien Son} 
\affiliation{Department of Physics, Chemistry and Biology, Link\"oping
  University, SE-581 83 Link\"oping, Sweden}
  
\author{Takeshi Ohshima}
\affiliation{National Institutes for Quantum and Radiological Science and Technology, 1233 Watanuki, Takasaki, Gunma 370-1292, Japan}

\author{Igor A. Abrikosov}
\affiliation{Department of Physics, Chemistry and Biology, Link\"oping
  University, SE-581 83 Link\"oping, Sweden}
\affiliation{Materials Modeling and Development Laboratory, National University of Science and Technology `MISIS', 119049 Moscow, Russia}

\author{Adam Gali} 
\email{gali.adam@wigner.mta.hu}
\affiliation{Wigner Research Centre for Physics, Hungarian Academy of Sciences,
  PO Box 49, H-1525, Budapest, Hungary}
\affiliation{Department of Atomic Physics, Budapest University of
  Technology and Economics, Budafoki \'ut 8., H-1111 Budapest,
  Hungary}

\date{\today}


\begin{abstract}
Identification of microscopic configuration of point defects acting as quantum bits is a key step in the advance  of quantum information processing and sensing. Among the numerous candidates, silicon vacancy related centers in silicon carbide (SiC) have shown remarkable properties owing to their particular spin-3/2 ground and excited states. Although, these centers were observed decades ago, still two competing models, the isolated negatively charged silicon vacancy and the complex of negatively charged silicon vacancy and neutral carbon vacancy [Phys. Rev. Lett.\ \textbf{115}, 247602 (2015)] are argued as an origin. By means of high precision first principles calculations and high resolution electron spin resonance measurements, we here unambiguously identify the Si-vacancy related qubits in hexagonal SiC as isolated negatively charged silicon vacancies. Moreover, we identify the Si-vacancy qubit configurations that provide room temperature optical readout. 
\end{abstract}
\maketitle



Point defects in solids acting as quantum bits (qubits) are highly promising platform for quantum information processing (QIP) and nanoscale sensor applications where typically their electron spin provides the functional quantum states. There are qubits that have long electron spin coherence times~\cite{Balasubramanian:NatMat2009, Koehl11, Christle2014, Widmann2014, Rose2017}, and some of them demonstrated to persist up to room temperature~\cite{Balasubramanian:NatMat2009, Widmann2014}. These electron spins can be optically initialized and readout~\cite{Jelezko:PSSa2006, Awschalom:Nature2010, Robledo:Nature2011, Koehl11, Falk2013}, making them very attractive candidates for QIP and related applications~\cite{Baranov2005, Gali11pss, Weber10}. Among these qubits, silicon-vacancy related defects in hexagonal polytypes of SiC, such as $4H$ and $6H$-SiC, have shown favorable spin properties \cite{Riedel2012, Kraus2014}, demonstrated even at single defect level at room temperature~\cite{Widmann2014}. Two and three different silicon vacancy related centers were observed  in $4H$ and $6H$-SiC, where the corresponding photoluminescence (PL) lines are denoted as V1 and V2 and V1, V2, and V3,~\cite{Sorman2000,Wagner2000} , respectively. V2 line in $4H$-SiC~\cite{Widmann2014} and V2 and V3 lines in $6H$-SiC~\cite{Riedel2012} are sufficiently strong to observe their corresponding electron spin via optically detected magnetic resonance (ODMR) measurements at room temperature. In particular, it has been demonstrated that V2 color center in $4H$-SiC can be used for magnetometer~\cite{Lee2015, Simin2016, Niethammer2016, Cochrane2016} and nano-scale thermometer~\cite{Anisimov2016} applications and as a room temperature maser~\cite{Kraus2014}.

\begin{figure}[h!]
	\includegraphics[width=0.95\columnwidth]{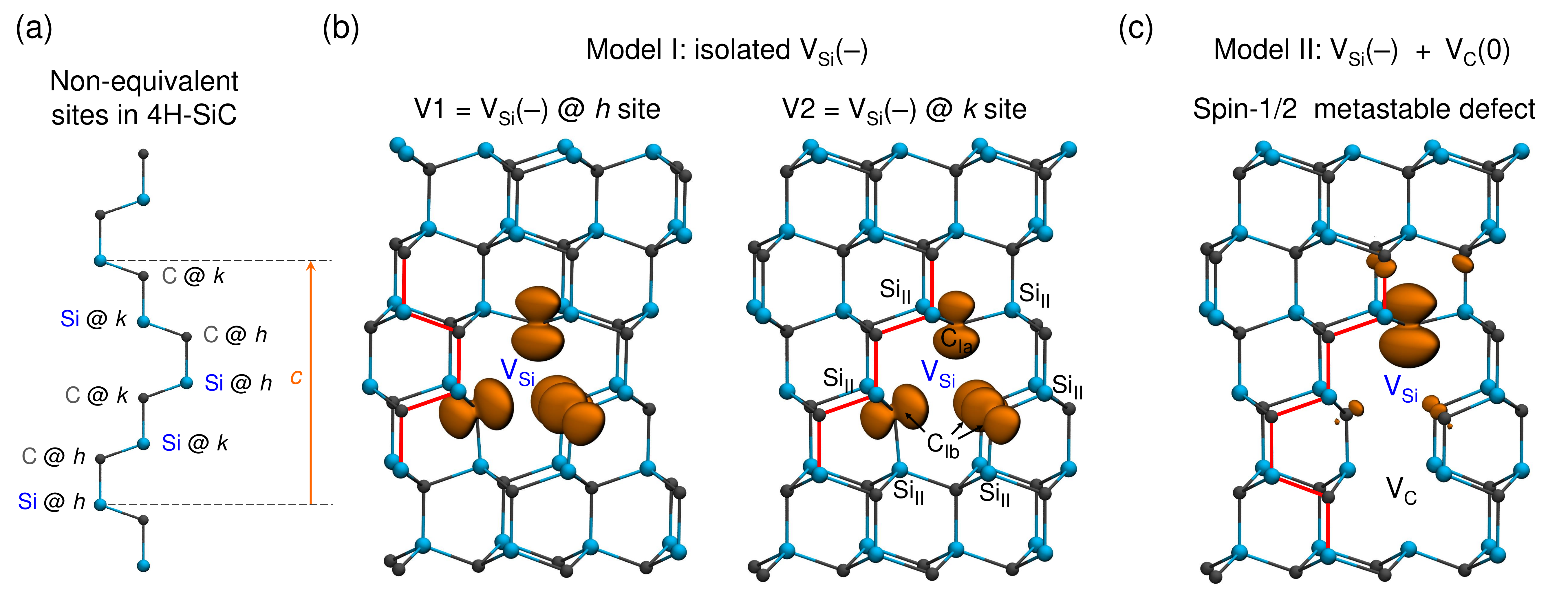}
	\caption{Models of V1-V2 silicon vacancy related qubits in $4H$-SiC. (a) Non-equivalent atomic sites, i.e. a quasi-hexagonal $h$ and a quasi-cubic $k$ sites, in the primitive cell of $4H$-SiC. (b) Isolated vacancy model and the assignment of V1-V2 centers to the different silicon vacancy configurations in $4H$-SiC. (c) Vacancy pair model of V2 center \cite{Kraus2014}. In (b) and (c) red lines highlights the stacking of the Si-C double layers to help to identify different configurations of silicon and carbon vacancies. Orange lobes show the spin density of the defects. Based on our first principles results, the isolated vacancy model in (b) can be assigned to V1 and V2 centers in $4H$-SiC. } 
	\label{fig:main}  
\end{figure}

Today, it is widely accepted that V1-V3 PL lines and Tv1-Tv3 electron paramagnetic resonance (EPR) signals in $4H$ and $6H$-SiC are related to spin-3/2 negatively charged silicon vacancies~\cite{Wimbauer97, Mizuochi2002, Orlinski2003, Mizuochi2005}. On the other hand, the actual microscopic configuration of these vacancy related centers is still debated. The unanswered question is whether these centers are isolated silicon vacancies (V$_{\text{Si}}(-)$ as model~I)~\cite{Mizuochi2002, Mizuochi2005, Janzen2009} or axial symmetric defect pairs, including a negatively charged silicon vacancy and a proximate neutral carbon vacancy~\cite{Kraus2014, Soltamov2015}  (V$_{\text{Si}}(-) + \text{V}_{\text{C}}(0)$ as model~II), see Fig.~\ref{fig:main}. 
An important difference of the two models is how the observed finite zero-field-splitting (ZFS) of the ground state spin sublevels is explained. In model~I, it is assumed that the C$_{3v}$ symmetric crystal field, allowing non-zero ZFS~\cite{Janzen2009}, is strong enough to cause a finite ZFS that accounts for the observations. Recent theoretical estimate on the ZFS of V2 center in $4H$-SiC supports this assumption~\cite{Soykal2016}. We note that Mizuochi and co-workers~\cite{Mizuochi2003} associated V2 center particularly with V$_{\text{Si}}(-)$ at $h$-site by comparing the similarities of V1-V2 and V1-V3 signals in $4H$ and $6H$-SiC, respectively.  Model~II, on the other hand, uses the non-distorted silicon vacancy model~\cite{Wimbauer97, Kraus2014}, where close to T$_d$ symmetry with negligible ZFS is assumed for an isolated V$_{\text{Si}}(-)$ in hexagonal SiC. In this model a proximate neutral carbon vacancy in a symmetrical configuration is assumed, see for example the model of V2 center in $4H$-SiC~\cite{Kraus2014} in Fig.~\ref{fig:main}(c), that lowers the symmetry of the silicon vacancy thus causing a finite ZFS. In a recent experiment on rhombic 15R-SiC~\cite{Soltamov2015}, silicon vacancy related centers were reported with similar~ characteristics as those in hexagonal SiC. Using electron nuclear double resonance (ENDOR) measurement, negative $^{29}$Si hyperfine coupling constants, i.e.\ negative electron spin density was observed for V2 center in 15R-SiC. As a proof of model II, this observation was attributed to the hyperfine coupling of the weakly negatively polarized silicon nuclei around the carbon vacancy~\cite{Soltamov2015}.  Identification of microscopic structure of the V1-2 centers in $4H$ SiC, i.e.\ validating one of these models, is an essential need for appropriate theoretical description and for controlled single defect fabrication purposes.
  
In this Rapid Communication, we show by means of high precision first principles calculations that the silicon vacancy-carbon vacancy pair model of V2 center in $4H$-SiC is a meta-stable configuration that has spin-1/2 ground state without any zero-field splitting and a hyper-fine signature that differs significantly from the experiment. Furthermore, we demonstrate by theoretical simulations and high resolution EPR measurements that the isolated silicon vacancy model accounts for majority of the observed magneto-optical properties of V1-V2 centers in $4H$-SiC. Especially, the simulated zero-phonon-line (ZPL) energies, the non-zero ZFS values, and the hyperfine structure that includes $^{29}$Si hyperfine values correspond to negative electron spin polarization are all in good agreement with the observations. Based on these results, we identify V1-V2 centers in $4H$-SiC as isolated negatively charged silicon vacancies. Furthermore, we identify the room temperature V2 Si-vacancy qubit at the $k$-site in $4H$-SiC, in contrast to previous assignments.

In our first principles point defect characterization study, we apply density functional theory (DFT) and supercell method to model single point defects. We apply plane wave basis set of 420~eV and standard projector-augmented wave~\cite{PAW} potentials as implemented in VASP code~\cite{VASP,VASP2}. For ZPL and hyperfine tensor calculations we use HSE06~\cite{HSE03,HSE06} hybrid exchange-correlation functional that has already demonstrated its predictive power for optical~\cite{Gali2009, Deak2010} and hyperfine properties~\cite{Szasz2013}. In the zero-field-splitting calculations we use PBE~\cite{PBE} functional that provides accurate results for defects in wide band gap semiconductors.~\cite{Ivady2014} According to previous theoretical ZFS studies on the NV center in diamond~\cite{Ivady2014, Falk2014} and divacancy in SiC~\cite{Falk2014} and our present results, the theoretical ZFS values have $\sim$16~MHz mean absolute error when compared with the experiment. Nevertheless, the ZFS values sensitively depend on the fine details of the crystal filed, thus tendencies observed in the calculated values can still be used for the identification of symmetrically non-equivalent  configurations of point defects, see for example Ref.~\onlinecite{Falk2014}.  For the sake of high numerical accuracy~\cite{Joel2017}, we employ 1532-atom $4H$-SiC supercell with $\Gamma$-point sampling of the Brillouin-zone. For structural optimization with HSE06 functional in these large supercells, the plane wave cut-off energy is slightly reduced to 390~eV, and a force criterion of 0.01~eV/\AA\ is applied. 
As the supercell size, $\approx$30~\AA\ in $c$ direction, is sufficiently large to properly accommodate both the isolated vacancy (model~I) and the defect pair (model~II), the results of these calculations are comparable.  

High precision EPR measurements in $4H$-SiC sample were performed on a Bruker X-band EPR spectrometer. High-purity semi-insulating (HPSI) bulk sample with large size was irradiated by 2~MeV electrons to a fluence of $8 \times 10^{18}$~cm$^{-2}$ at room temperature and annealed at $\sim$400 $^{\circ}$C in order to remove the interference of other EPR centers related to interstitial defects. EPR measurements were performed in darkness at room temperature. For further details on EPR experiments in $4H$ and $6H$-SiC see Ref.~[\onlinecite{Note1}]. 

\begin{figure}[h!]
	\includegraphics[width=0.80\columnwidth]{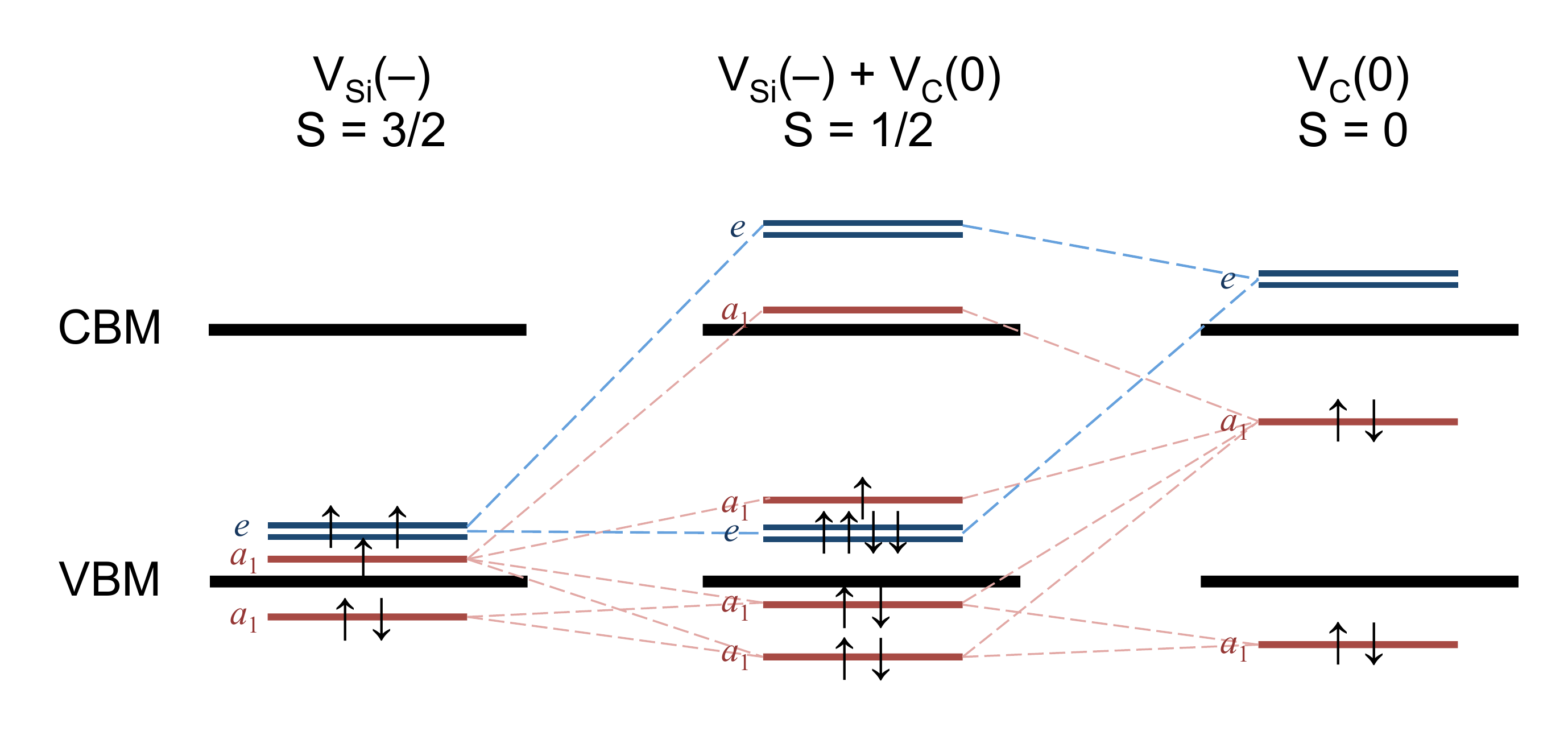}
	\caption{ Electronic structure of  $\text{V}_{\text{Si}}(-) + \text{V}_{\text{C}}(0)$ complex: defect molecule diagram analysis with the states of the isolated silicon and carbon vacancies. Positions of one-particle states for $\text{V}_{\text{Si}}(-)$ and $\text{V}_{\text{Si}}(-) + \text{V}_{\text{C}}(0)$ reflect the results of our \emph{ab initio} calculations. For the electronic structure of carbon vacancy in $4H$-SiC see Ref~\cite{Trinh2013}.} 
	\label{fig:elstruct}  
\end{figure}

First, we carry out high precision first principles calculation on model~II. We consider the nearest V$_{\text{Si}}(-) + \text{V}_{\text{C}}(0)$ pair not sharing the same Si-C bilayer [see Fig.~\ref{fig:main}(c)], which is the suggested configuration of V2 center in $4H$, $6H$, and $15R$-SiC~\cite{Kraus2014,Soltamov2015}. In the simulations, we observe notable interaction between the vacancies that results in a weakly bonded defect pair with the electronic structure that significantly differs from the electronic structure of the isolated silicon vacancy, see Fig.~\ref{fig:elstruct}. In tight binding picture, both isolated V$_{\text{Si}}(-)$ and isolated V$_{\text{C}}(0)$ possess two non-degenerate $a_1$ states and a double degenerate $e$ state. In the case of isolated V$_{\text{Si}}(-)$ an $a_1$ state and an $e$ state appear in the band gap with increasing energy. In the  V$_{\text{Si}}(-) + \text{V}_{\text{C}}(0)$ pair, the $e$ states of the vacancies form a weakly bounding $e$ state that falls into the band gap and is mainly localized on the silicon vacancy site. Beside this state, an $a_1$ state of anti-bonding nature appears into the band gap slightly above the $e$ state. Altogether, five electrons can be found in the band gap that fully occupy the lower lying $e$ state and partially occupy the $a_1$ state. The ground state of the $\text{V}_{\text{Si}}(-) + \text{V}_{\text{C}}(0)$ pair defect is thus spin-1/2, see the spin density in Fig.~\ref{fig:main}(c). The hyperfine signature of the defect substantially deviates from the V1-V2 centers' hyperfine signature in $4H$ SiC \cite{Note1}. Furthermore, by comparing the formation energy of model~II and the negatively charged divacancy (immediate neighbor vacancies sharing the same Si-C bilayer), we find that the latter is lower in energy by 1.58~eV, showing that model~II is a meta-stable configuration of the negatively charged divacancy, which presumably anneals out at the temperatures where carbon vacancies are mobile~\cite{Son2007}. Based on these results we argue that the silicon vacancy-carbon vacancy pair model is not appropriate for the silicon vacancy related centers in SiC.

\begin{figure}[h!]
	\includegraphics[width=0.9\columnwidth]{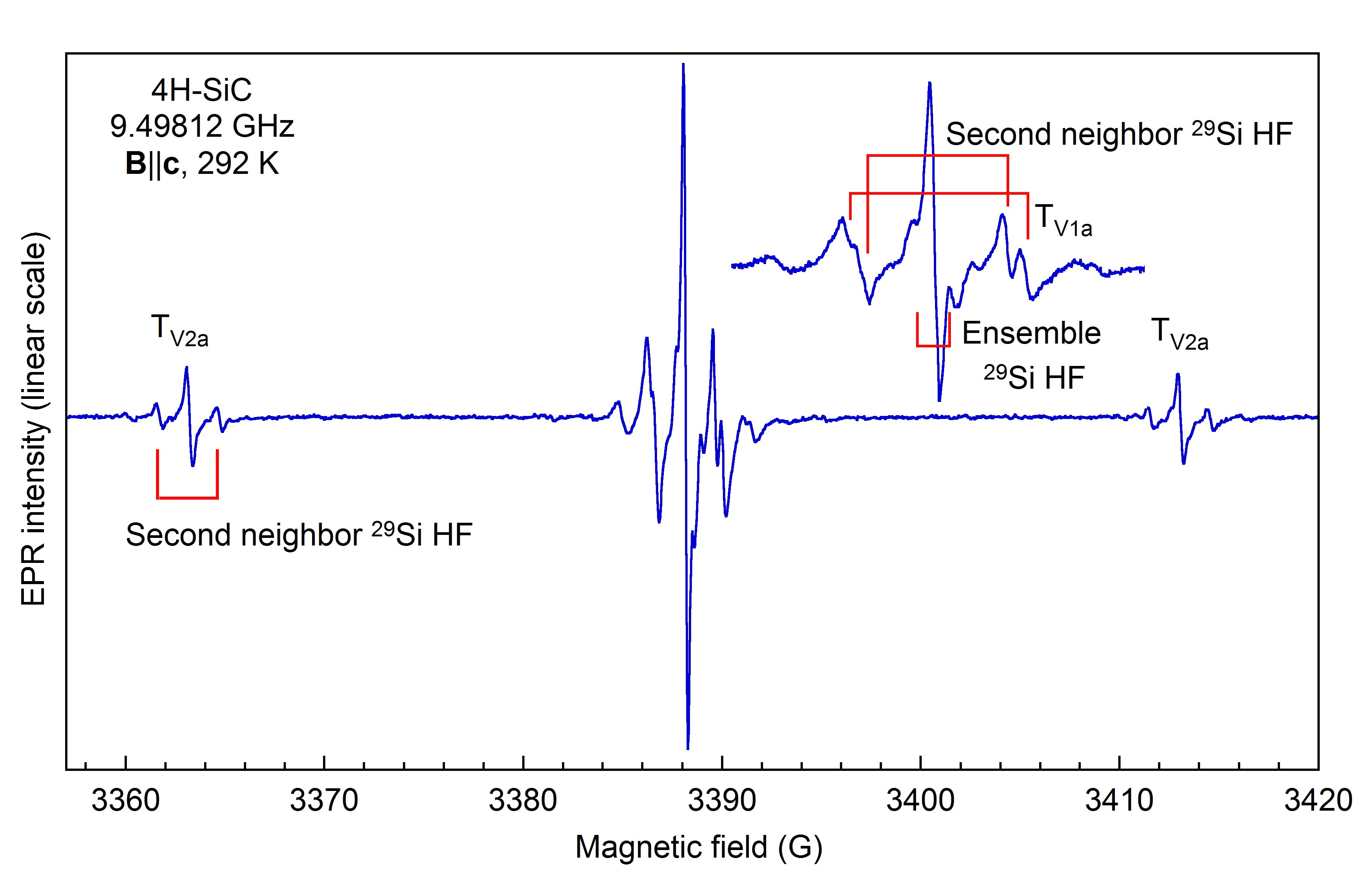}
	\caption{  EPR spectra of the T$_{\text{V1a}}$ and T$_{\text{V2a}}$ centers in $4H$-SiC measured in darkness at 292~K for $B\parallel c$ using low microwave power (MW) of 2~$\mu$W and a low modulation field of 0.01~G. In an extended magnetic field scale, the inset shows the ZFS of T$_{\text{V1a}}$, which partly overlaps with the Si hyperfine structure, and a small Si hyperfine splitting of $\sim$0.8~G ($\sim$2.2~MHz). } 
	\label{fig:EPR1}  
\end{figure}

We further note that the negative anisotropic $^{29}$Si hyperfine splitting in the range of 1.3-2.2~MHz observed by ENDOR~\cite{Soltamov2015} can be explained by isolated  V$_{\text{Si}}(-)$ and it is not an evidence of the presence of carbon vacancy. As spin density is the fingerprint of a complex many-body wavefunction, sign changes in the spin density may occur even for a simple point defect, for instance, the NV center diamond~\cite{Gali2009}. As our DFT simulations can capture these effects, we investigate hyperfine interactions of both $^{29}$Si and $^{13}$C nuclei up to 7.8~\AA\ distance from the isolated silicon vacancy (model I)~\cite{Note1}. We find negative anisotropic $^{29}$Si hyperfine splitting in the order of few MHz  for both silicon vacancy configurations in $4H$-SiC. In particular, $A_z = -1.3$ -- $ -2.1$~MHz is found for several Si sites 6.1~\AA\ away from the silicon vacancy and $A_z = -5.1$~MHz is found for one Si site 5.0~\AA\ away from $\text{V}_{\text{Si}}$ at $k$ site~\cite{Note1}.

Fig.~\ref{fig:EPR1} shows a high resolution EPR spectrum in 4H-SiC measured at 292 K in darkness for $B\parallel c$. With low MW and low field modulation, the overlapping between the ZFS components of the TV1a center (with a splitting of $\sim$3.66 G) and the hyperfine structure due to the interaction with the nuclear spin of one $^{29}$Si among 12 equivalent Si in the second neighbor ($\sim$3 G) can be resolved. The observed ZFS and the hyperfine constants of the interaction with 1 C and 3 C in the nearest neighbor and with 12 Si in the second neighbor for TV1a and TV2a centers in 4H-SiC are given in Table.~\ref{tab:4H}. Corresponding data for 6H-SiC can be found in Ref.~\cite{Note1}. Furthermore, a small hyperfine splitting of $\approx$0.8~G ($\approx$2.2~MHz) due to the hyperfine interaction with $^{29}$Si nuclei in further neighbor shells beyond the second neighbor is also observed in $4H$-SiC (see the inset in Fig.~\ref{fig:EPR1}). Note that our first principles calculation on the isolated silicon vacancy in $4H$-SiC can account for this splitting, see the negative hyperfine coupling constants above and Ref.~\cite{Note1}. Furthermore, the magnitude of the negative hypefine coupling constants reported for the silicon vacancey related centers in 15R-SiC~\cite{Soltamov2015} is similar to the $\approx$0.8~G ($\approx$2.2~MHz) splitting observed in our EPR results in $4H$-SiC.  Since the electronic structure of the isolated vacancy in $15R$-SiC is akin to their counterparts' electronic structure in $4H$-SiC,  similar negative spin density shell, resulting in negative $^{29}$Si hyperfine constants, should form around the silicon vacancies in $15R$-SiC. Our results indicate that the isolated V$_{\text{Si}}(-)$ can account for the ENDOR signatures recorded in $15R$-SiC fairly well. 

\begin{table}[!h]
\caption{Theoretical and experimental magneto-optical data for V1 and V2 centers in $4H$-SiC. The hyperfine splitting at $B\parallel c$ was determined for the first neighbor $^{13}$C nuclei on ($1\times \text{C}_{\text{Ia}}$) and out ($3\times \text{C}_{\text{Ib}}$) of the symmetry axis of the isolated silicon vacancy  and the averaged hyperfine splitting for the twelve second neighbor $^{29}$Si sites ($12\times \tilde{\text{Si}}_{\text{II}}$). See the considered nuclei sites in Fig.\ref{fig:main}(b). Experimental ZPL energies were reported in Ref.~\onlinecite{Wagner2000} while the ZFS and the resolvable hyperfine values are determined by our EPR measurements, see Fig.~\ref{fig:EPR1}. The good agreement between theory and experiment supporting the isolated silicon vacancy model of V1-V2 centers.}
\begin{ruledtabular}
       \begin{tabular}{c|c|c|ccc} 
             Center /       &   \multirow{2}{*}{ ZFS (MHz) }  & \multirow{2}{*}{ZPL (eV)} & \multicolumn{3}{c}{Hypefine splitting (MHz)}      \\
             Configuration &                                               &                                       &  $1 \times \text{C}_{\text{Ia}}$  &  $3 \times \text{C}_{\text{Ib}}$   &  $12 \times \tilde{\text{Si}}_{\text{II}}$  \\  \hline \hline
             \multicolumn{6}{c}{Experiment}
             \\ \hline
           V1  &   2.6     & 1.438   & 79.9  & 39.2  & 8.2 \\
           V2  &   35.0   &  1.352  & 80.4  & 37.0  & 8.4  \\ \hline
            \multicolumn{6}{c}{Theory} \\ \hline
           V$_{\text{Si}}^{-}$@$h$   &   18.3   &  1.541  & 85.5 & 40.6 & 7.7 \\
           V$_{\text{Si}}^{-}$@$k$   &   33.3   &  1.443  & 84.7 & 38.7 & 7.9 \\
       \end{tabular}
\end{ruledtabular}\label{tab:4H}
\end{table}

Next, we thoroughly investigate the isolated V$_{\text{Si}}(-)$ model in $4H$-SiC. The multiplet structure of these defects includes a $^{4}$A$_2$ groundstate, a low energy $^{4}$A$_2$ and $^{4}$E optically allowed excited states, and other spin-1/2 shelving states between the ground and optical excited states~\cite{Soykal2016}. Accordingly, in our first principles calculations, we consider the $^{4}$A$_2$ ground state and the lowest energy optically excited state, either the $^{4}$A$_2$ or the $^{4}$E state. We calculate ground state ZFS values, ZPL line energies, and groundstate hyperfine splitting for $B \parallel c$ and compare them with existing ZPL data and with the results of our EPR ZFS and hyperfine splitting measurements. Our theoretical and experimental ZFS values are given in Table~\ref{tab:4H}. Importantly, the theoretically predicted ZFS is non-zero for all the symmetrically non-equivalent vacancy configurations. In agreement with previous theoretical estimates \cite{Soykal2016}, these results disprove the existence of undistorted $T_d$ symmetric silicon vacancy~\cite{Wimbauer97} in $4H$-SiC, which was one of the basic assumption of model~II~\cite{Kraus2014}. Furthermore, both the calculated ZPL energies and the hyperfine values are in good agreement with the experimental data. 

Finally, since not all of the silicon vacancy centers have room temperature ODMR signal, they have different potential for qubit applications. The V2 center in $4H$-SiC exhibits observable ODMR signal even at room temperature and its spin ensemble has been used for magnetometry~\cite{Lee2015, Simin2016, Niethammer2016, Cochrane2016}. Thus with improving the brightness and ODMR contrast of these single emitters is promising for room temperature nanoscale magnetometry of biological molecules and quantum information processing applications~\cite{Baranov2005, Widmann2014}. For proper theoretical description and single defect engineering, the actual configuration of this center should be determined.  Therefore, here we assign the symmetrically non-equivalent isolated silicon vacancy configurations, see Fig.~\ref{fig:main}(a), to the V1-V2 centers in $4H$-SiC. Note that such assignment requires especially high precision from both the theoretical and the experimental results. As can be seen in Table~\ref{tab:4H}, the ZFS results, the ZPL energies, and the vast majority of hyperfine constants in the first and second shell around the vacancy~\cite{Note2} support the identification of V1 center as V$_{\text{Si}}(-)$ at $h$ site and V2 center as V$_{\text{Si}}(-)$ at $k$ site. Accordingly, we assign V1 and V2 centers in $4H$-SiC to $h$ and $k$ configuration of V$_{\text{Si}}(-)$, respectively. Note that this assignment is in contrast to the previously suggested identification~\cite{Mizuochi2003, Wagner2000} that relies on the comparison of $4H$ and $6H$-SiC magneto-optical spectra. This approach, however, can be misleading for defect configuration identification in hexagonal SiC~\cite{Szasz2014}. 

In summary, we investigated the microscopic origin of V1-V2 centers in $4H$-SiC.  We demonstrated by first principles calculations that the silicon vacancy - carbon vacancy defect is metastable and possesses $S=1/2$ ground state, in stark contrast to the properties of the V1-V2 centers. We showed, however, that the isolated negatively charged silicon vacancy can accurately reproduce the reported magneto-optical data of these centers, including the hyperfine signatures of $^{29}$Si nuclear spins. Furthermore, we identified the room temperature V2 qubit as V$_{\text{Si}}(-)$ at the $k$ site in $4H$-SiC.

\section*{Acknowledgments} 

Support from the Knut \& Alice Wallenberg Foundation project Strong Field Physics and New States of Matter 2014-2019 (COTXS), the Grant of the Ministry of Education and Science of the Russian Federation  (grant No.\ 14.Y26.31.0005),  the Swedish Research Council (VR 2016-04068), the Carl-Trygger Stiftelse f\"or Vetenskaplig Forskning (CTS 15:339), JSPS KAKENHI B 26286047, and the Hungarian NKFIH Grant No.\ NVKP\_16-1-2016-0152958 is acknowledged. The calculations were performed on resources provided by the Swedish National Infrastructure for Computing (SNIC 2016/1-528) at the National Supercomputer Centre (NSC)  and by Link\"oping University (LiU-2015-00017-60).


\begin{thebibliography}{46}%
\makeatletter
\providecommand \@ifxundefined [1]{%
 \@ifx{#1\undefined}
}%
\providecommand \@ifnum [1]{%
 \ifnum #1\expandafter \@firstoftwo
 \else \expandafter \@secondoftwo
 \fi
}%
\providecommand \@ifx [1]{%
 \ifx #1\expandafter \@firstoftwo
 \else \expandafter \@secondoftwo
 \fi
}%
\providecommand \natexlab [1]{#1}%
\providecommand \enquote  [1]{``#1''}%
\providecommand \bibnamefont  [1]{#1}%
\providecommand \bibfnamefont [1]{#1}%
\providecommand \citenamefont [1]{#1}%
\providecommand \href@noop [0]{\@secondoftwo}%
\providecommand \href [0]{\begingroup \@sanitize@url \@href}%
\providecommand \@href[1]{\@@startlink{#1}\@@href}%
\providecommand \@@href[1]{\endgroup#1\@@endlink}%
\providecommand \@sanitize@url [0]{\catcode `\\12\catcode `\$12\catcode
  `\&12\catcode `\#12\catcode `\^12\catcode `\_12\catcode `\%12\relax}%
\providecommand \@@startlink[1]{}%
\providecommand \@@endlink[0]{}%
\providecommand \url  [0]{\begingroup\@sanitize@url \@url }%
\providecommand \@url [1]{\endgroup\@href {#1}{\urlprefix }}%
\providecommand \urlprefix  [0]{URL }%
\providecommand \Eprint [0]{\href }%
\providecommand \doibase [0]{http://dx.doi.org/}%
\providecommand \selectlanguage [0]{\@gobble}%
\providecommand \bibinfo  [0]{\@secondoftwo}%
\providecommand \bibfield  [0]{\@secondoftwo}%
\providecommand \translation [1]{[#1]}%
\providecommand \BibitemOpen [0]{}%
\providecommand \bibitemStop [0]{}%
\providecommand \bibitemNoStop [0]{.\EOS\space}%
\providecommand \EOS [0]{\spacefactor3000\relax}%
\providecommand \BibitemShut  [1]{\csname bibitem#1\endcsname}%
\let\auto@bib@innerbib\@empty
\bibitem [{\citenamefont {Balasubramanian}\ \emph {et~al.}(2009)\citenamefont
  {Balasubramanian}, \citenamefont {Neumann}, \citenamefont {Twitchen},
  \citenamefont {Markham}, \citenamefont {Kolesov}, \citenamefont {Mizuochi},
  \citenamefont {Isoya}, \citenamefont {Achard}, \citenamefont {Beck},
  \citenamefont {Tissler}, \citenamefont {Jacques}, \citenamefont {Hemmer},
  \citenamefont {Jelezko},\ and\ \citenamefont
  {Wrachtrup}}]{Balasubramanian:NatMat2009}%
  \BibitemOpen
  \bibfield  {author} {\bibinfo {author} {\bibfnamefont {G.}~\bibnamefont
  {Balasubramanian}}, \bibinfo {author} {\bibfnamefont {P.}~\bibnamefont
  {Neumann}}, \bibinfo {author} {\bibfnamefont {D.}~\bibnamefont {Twitchen}},
  \bibinfo {author} {\bibfnamefont {M.}~\bibnamefont {Markham}}, \bibinfo
  {author} {\bibfnamefont {R.}~\bibnamefont {Kolesov}}, \bibinfo {author}
  {\bibfnamefont {N.}~\bibnamefont {Mizuochi}}, \bibinfo {author}
  {\bibfnamefont {J.}~\bibnamefont {Isoya}}, \bibinfo {author} {\bibfnamefont
  {J.}~\bibnamefont {Achard}}, \bibinfo {author} {\bibfnamefont
  {J.}~\bibnamefont {Beck}}, \bibinfo {author} {\bibfnamefont {J.}~\bibnamefont
  {Tissler}}, \bibinfo {author} {\bibfnamefont {V.}~\bibnamefont {Jacques}},
  \bibinfo {author} {\bibfnamefont {P.~R.}\ \bibnamefont {Hemmer}}, \bibinfo
  {author} {\bibfnamefont {F.}~\bibnamefont {Jelezko}}, \ and\ \bibinfo
  {author} {\bibfnamefont {J.}~\bibnamefont {Wrachtrup}},\ }\href
  {http://dx.doi.org/10.1038/nmat2420} {\bibfield  {journal} {\bibinfo
  {journal} {Nat. Mater.}\ }\textbf {\bibinfo {volume} {8}},\ \bibinfo {pages}
  {383} (\bibinfo {year} {2009})}\BibitemShut {NoStop}%
\bibitem [{\citenamefont {Koehl}\ \emph {et~al.}(2011)\citenamefont {Koehl},
  \citenamefont {Buckley}, \citenamefont {Heremans}, \citenamefont {Calusine},\
  and\ \citenamefont {Awschalom}}]{Koehl11}%
  \BibitemOpen
  \bibfield  {author} {\bibinfo {author} {\bibfnamefont {W.~F.}\ \bibnamefont
  {Koehl}}, \bibinfo {author} {\bibfnamefont {B.~B.}\ \bibnamefont {Buckley}},
  \bibinfo {author} {\bibfnamefont {F.~J.}\ \bibnamefont {Heremans}}, \bibinfo
  {author} {\bibfnamefont {G.}~\bibnamefont {Calusine}}, \ and\ \bibinfo
  {author} {\bibfnamefont {D.~D.}\ \bibnamefont {Awschalom}},\ }\href@noop {}
  {\bibfield  {journal} {\bibinfo  {journal} {Nature}\ }\textbf {\bibinfo
  {volume} {479}},\ \bibinfo {pages} {84} (\bibinfo {year} {2011})}\BibitemShut
  {NoStop}%
\bibitem [{\citenamefont {Christle}\ \emph {et~al.}(2015)\citenamefont
  {Christle}, \citenamefont {Falk}, \citenamefont {Andrich}, \citenamefont
  {Klimov}, \citenamefont {Hassan}, \citenamefont {Son}, \citenamefont
  {Janz{\'e}n}, \citenamefont {Ohshima},\ and\ \citenamefont
  {Awschalom}}]{Christle2014}%
  \BibitemOpen
  \bibfield  {author} {\bibinfo {author} {\bibfnamefont {D.~J.}\ \bibnamefont
  {Christle}}, \bibinfo {author} {\bibfnamefont {A.~L.}\ \bibnamefont {Falk}},
  \bibinfo {author} {\bibfnamefont {P.}~\bibnamefont {Andrich}}, \bibinfo
  {author} {\bibfnamefont {P.~V.}\ \bibnamefont {Klimov}}, \bibinfo {author}
  {\bibfnamefont {J.~U.}\ \bibnamefont {Hassan}}, \bibinfo {author}
  {\bibfnamefont {N.~T.}\ \bibnamefont {Son}}, \bibinfo {author} {\bibfnamefont
  {E.}~\bibnamefont {Janz{\'e}n}}, \bibinfo {author} {\bibfnamefont
  {T.}~\bibnamefont {Ohshima}}, \ and\ \bibinfo {author} {\bibfnamefont
  {D.~D.}\ \bibnamefont {Awschalom}},\ }\href
  {http://dx.doi.org/10.1038/nmat4144} {\bibfield  {journal} {\bibinfo
  {journal} {Nat. Mater.}\ }\textbf {\bibinfo {volume} {14}},\ \bibinfo {pages}
  {160} (\bibinfo {year} {2015})}\BibitemShut {NoStop}%
\bibitem [{\citenamefont {Widmann}\ \emph {et~al.}(2015)\citenamefont
  {Widmann}, \citenamefont {Lee}, \citenamefont {Rendler}, \citenamefont {Son},
  \citenamefont {Fedder}, \citenamefont {Paik}, \citenamefont {Yang},
  \citenamefont {Zhao}, \citenamefont {Yang}, \citenamefont {Booker},
  \citenamefont {Denisenko}, \citenamefont {Jamali}, \citenamefont
  {Momenzadeh}, \citenamefont {Gerhardt}, \citenamefont {Ohshima},
  \citenamefont {Gali}, \citenamefont {Janz{\'e}n},\ and\ \citenamefont
  {Wrachtrup}}]{Widmann2014}%
  \BibitemOpen
  \bibfield  {author} {\bibinfo {author} {\bibfnamefont {M.}~\bibnamefont
  {Widmann}}, \bibinfo {author} {\bibfnamefont {S.-Y.}\ \bibnamefont {Lee}},
  \bibinfo {author} {\bibfnamefont {T.}~\bibnamefont {Rendler}}, \bibinfo
  {author} {\bibfnamefont {N.~T.}\ \bibnamefont {Son}}, \bibinfo {author}
  {\bibfnamefont {H.}~\bibnamefont {Fedder}}, \bibinfo {author} {\bibfnamefont
  {S.}~\bibnamefont {Paik}}, \bibinfo {author} {\bibfnamefont {L.-P.}\
  \bibnamefont {Yang}}, \bibinfo {author} {\bibfnamefont {N.}~\bibnamefont
  {Zhao}}, \bibinfo {author} {\bibfnamefont {S.}~\bibnamefont {Yang}}, \bibinfo
  {author} {\bibfnamefont {I.}~\bibnamefont {Booker}}, \bibinfo {author}
  {\bibfnamefont {A.}~\bibnamefont {Denisenko}}, \bibinfo {author}
  {\bibfnamefont {M.}~\bibnamefont {Jamali}}, \bibinfo {author} {\bibfnamefont
  {S.~A.}\ \bibnamefont {Momenzadeh}}, \bibinfo {author} {\bibfnamefont
  {I.}~\bibnamefont {Gerhardt}}, \bibinfo {author} {\bibfnamefont
  {T.}~\bibnamefont {Ohshima}}, \bibinfo {author} {\bibfnamefont
  {A.}~\bibnamefont {Gali}}, \bibinfo {author} {\bibfnamefont {E.}~\bibnamefont
  {Janz{\'e}n}}, \ and\ \bibinfo {author} {\bibfnamefont {J.}~\bibnamefont
  {Wrachtrup}},\ }\href {http://dx.doi.org/10.1038/nmat4145} {\bibfield
  {journal} {\bibinfo  {journal} {Nat. Mater.}\ }\textbf {\bibinfo {volume}
  {14}},\ \bibinfo {pages} {164} (\bibinfo {year} {2015})}\BibitemShut
  {NoStop}%
\bibitem [{\citenamefont {{Rose}}\ \emph {et~al.}(2017)\citenamefont {{Rose}},
  \citenamefont {{Huang}}, \citenamefont {{Zhang}}, \citenamefont
  {{Tyryshkin}}, \citenamefont {{Sangtawesin}}, \citenamefont {{Srinivasan}},
  \citenamefont {{Loudin}}, \citenamefont {{Markham}}, \citenamefont
  {{Edmonds}}, \citenamefont {{Twitchen}}, \citenamefont {{Lyon}},\ and\
  \citenamefont {{de Leon}}}]{Rose2017}%
  \BibitemOpen
  \bibfield  {author} {\bibinfo {author} {\bibfnamefont {B.~C.}\ \bibnamefont
  {{Rose}}}, \bibinfo {author} {\bibfnamefont {D.}~\bibnamefont {{Huang}}},
  \bibinfo {author} {\bibfnamefont {Z.-H.}\ \bibnamefont {{Zhang}}}, \bibinfo
  {author} {\bibfnamefont {A.~M.}\ \bibnamefont {{Tyryshkin}}}, \bibinfo
  {author} {\bibfnamefont {S.}~\bibnamefont {{Sangtawesin}}}, \bibinfo {author}
  {\bibfnamefont {S.}~\bibnamefont {{Srinivasan}}}, \bibinfo {author}
  {\bibfnamefont {L.}~\bibnamefont {{Loudin}}}, \bibinfo {author}
  {\bibfnamefont {M.~L.}\ \bibnamefont {{Markham}}}, \bibinfo {author}
  {\bibfnamefont {A.~M.}\ \bibnamefont {{Edmonds}}}, \bibinfo {author}
  {\bibfnamefont {D.~J.}\ \bibnamefont {{Twitchen}}}, \bibinfo {author}
  {\bibfnamefont {S.~A.}\ \bibnamefont {{Lyon}}}, \ and\ \bibinfo {author}
  {\bibfnamefont {N.~P.}\ \bibnamefont {{de Leon}}},\ }\href@noop {} {\bibfield
   {journal} {\bibinfo  {journal} {ArXiv e-prints}\ } (\bibinfo {year}
  {2017})},\ \Eprint {http://arxiv.org/abs/1706.01555} {arXiv:1706.01555
  [cond-mat.mtrl-sci]} \BibitemShut {NoStop}%
\bibitem [{\citenamefont {Jelezko}\ and\ \citenamefont
  {Wrachtrup}(2006)}]{Jelezko:PSSa2006}%
  \BibitemOpen
  \bibfield  {author} {\bibinfo {author} {\bibfnamefont {F.}~\bibnamefont
  {Jelezko}}\ and\ \bibinfo {author} {\bibfnamefont {J.}~\bibnamefont
  {Wrachtrup}},\ }\href {http://dx.doi.org/10.1002/pssa.200671403} {\bibfield
  {journal} {\bibinfo  {journal} {Phys. Stat. Sol. A}\ }\textbf {\bibinfo
  {volume} {203}},\ \bibinfo {pages} {3207} (\bibinfo {year}
  {2006})}\BibitemShut {NoStop}%
\bibitem [{\citenamefont {Buckley}\ \emph {et~al.}(2010)\citenamefont
  {Buckley}, \citenamefont {Fuchs}, \citenamefont {Bassett},\ and\
  \citenamefont {Awschalom}}]{Awschalom:Nature2010}%
  \BibitemOpen
  \bibfield  {author} {\bibinfo {author} {\bibfnamefont {B.~B.}\ \bibnamefont
  {Buckley}}, \bibinfo {author} {\bibfnamefont {G.~D.}\ \bibnamefont {Fuchs}},
  \bibinfo {author} {\bibfnamefont {L.~C.}\ \bibnamefont {Bassett}}, \ and\
  \bibinfo {author} {\bibfnamefont {D.~D.}\ \bibnamefont {Awschalom}},\ }\href
  {http://dx.doi.org/10.1126/science.1196436} {\bibfield  {journal} {\bibinfo
  {journal} {Science}\ }\textbf {\bibinfo {volume} {330}},\ \bibinfo {pages}
  {1212} (\bibinfo {year} {2010})}\BibitemShut {NoStop}%
\bibitem [{\citenamefont {Robledo}\ \emph {et~al.}(2011)\citenamefont
  {Robledo}, \citenamefont {Childress}, \citenamefont {Bernien}, \citenamefont
  {Hensen}, \citenamefont {Alkemade},\ and\ \citenamefont
  {Hanson}}]{Robledo:Nature2011}%
  \BibitemOpen
  \bibfield  {author} {\bibinfo {author} {\bibfnamefont {L.}~\bibnamefont
  {Robledo}}, \bibinfo {author} {\bibfnamefont {L.}~\bibnamefont {Childress}},
  \bibinfo {author} {\bibfnamefont {H.}~\bibnamefont {Bernien}}, \bibinfo
  {author} {\bibfnamefont {B.}~\bibnamefont {Hensen}}, \bibinfo {author}
  {\bibfnamefont {P.~F.~A.}\ \bibnamefont {Alkemade}}, \ and\ \bibinfo {author}
  {\bibfnamefont {R.}~\bibnamefont {Hanson}},\ }\href
  {http://dx.doi.org/10.1038/nature10401} {\bibfield  {journal} {\bibinfo
  {journal} {Nature}\ }\textbf {\bibinfo {volume} {477}},\ \bibinfo {pages}
  {574} (\bibinfo {year} {2011})}\BibitemShut {NoStop}%
\bibitem [{\citenamefont {Falk}\ \emph {et~al.}()\citenamefont {Falk},
  \citenamefont {Buckley}, \citenamefont {Calusine}, \citenamefont {Koehl},
  \citenamefont {Dobrovitski}, \citenamefont {Politi}, \citenamefont {Zorman},
  \citenamefont {Feng},\ and\ \citenamefont {Awschalom}}]{Falk2013}%
  \BibitemOpen
  \bibfield  {author} {\bibinfo {author} {\bibfnamefont {A.~L.}\ \bibnamefont
  {Falk}}, \bibinfo {author} {\bibfnamefont {B.~B.}\ \bibnamefont {Buckley}},
  \bibinfo {author} {\bibfnamefont {G.}~\bibnamefont {Calusine}}, \bibinfo
  {author} {\bibfnamefont {W.~F.}\ \bibnamefont {Koehl}}, \bibinfo {author}
  {\bibfnamefont {V.~V.}\ \bibnamefont {Dobrovitski}}, \bibinfo {author}
  {\bibfnamefont {A.}~\bibnamefont {Politi}}, \bibinfo {author} {\bibfnamefont
  {C.~A.}\ \bibnamefont {Zorman}}, \bibinfo {author} {\bibfnamefont {P.~X.-L.}\
  \bibnamefont {Feng}}, \ and\ \bibinfo {author} {\bibfnamefont {D.~D.}\
  \bibnamefont {Awschalom}},\ }\href@noop {} {\bibfield  {journal} {\bibinfo
  {journal} {Nat. Commun.}\ }\textbf {\bibinfo {volume} {4}},\ \bibinfo {pages}
  {1819}}\BibitemShut {NoStop}%
\bibitem [{\citenamefont {Baranov}\ \emph {et~al.}(2005)\citenamefont
  {Baranov}, \citenamefont {Il'in}, \citenamefont {Mokhov}, \citenamefont
  {Muzafarova}, \citenamefont {Orlinskii},\ and\ \citenamefont
  {Schmidt}}]{Baranov2005}%
  \BibitemOpen
  \bibfield  {author} {\bibinfo {author} {\bibfnamefont {P.}~\bibnamefont
  {Baranov}}, \bibinfo {author} {\bibfnamefont {I.}~\bibnamefont {Il'in}},
  \bibinfo {author} {\bibfnamefont {E.}~\bibnamefont {Mokhov}}, \bibinfo
  {author} {\bibfnamefont {M.}~\bibnamefont {Muzafarova}}, \bibinfo {author}
  {\bibfnamefont {S.}~\bibnamefont {Orlinskii}}, \ and\ \bibinfo {author}
  {\bibfnamefont {J.}~\bibnamefont {Schmidt}},\ }\href {\doibase
  10.1134/1.2142873} {\bibfield  {journal} {\bibinfo  {journal} {JETP Letters}\
  }\textbf {\bibinfo {volume} {82}},\ \bibinfo {pages} {441} (\bibinfo {year}
  {2005})}\BibitemShut {NoStop}%
\bibitem [{\citenamefont {Gali}(2011)}]{Gali11pss}%
  \BibitemOpen
  \bibfield  {author} {\bibinfo {author} {\bibfnamefont {A.}~\bibnamefont
  {Gali}},\ }\href {\doibase 10.1002/pssb.201046254} {\bibfield  {journal}
  {\bibinfo  {journal} {physica status solidi (b)}\ }\textbf {\bibinfo {volume}
  {248}},\ \bibinfo {pages} {1337} (\bibinfo {year} {2011})}\BibitemShut
  {NoStop}%
\bibitem [{\citenamefont {Weber}\ \emph {et~al.}(2010)\citenamefont {Weber},
  \citenamefont {Koehl}, \citenamefont {Varley}, \citenamefont {Janotti},
  \citenamefont {Buckley}, \citenamefont {Van~de Walle},\ and\ \citenamefont
  {Awschalom}}]{Weber10}%
  \BibitemOpen
  \bibfield  {author} {\bibinfo {author} {\bibfnamefont {J.~R.}\ \bibnamefont
  {Weber}}, \bibinfo {author} {\bibfnamefont {W.~F.}\ \bibnamefont {Koehl}},
  \bibinfo {author} {\bibfnamefont {J.~B.}\ \bibnamefont {Varley}}, \bibinfo
  {author} {\bibfnamefont {A.}~\bibnamefont {Janotti}}, \bibinfo {author}
  {\bibfnamefont {B.~B.}\ \bibnamefont {Buckley}}, \bibinfo {author}
  {\bibfnamefont {C.~G.}\ \bibnamefont {Van~de Walle}}, \ and\ \bibinfo
  {author} {\bibfnamefont {D.~D.}\ \bibnamefont {Awschalom}},\ }\href {\doibase
  10.1073/pnas.1003052107} {\bibfield  {journal} {\bibinfo  {journal} {PNAS}\
  }\textbf {\bibinfo {volume} {107}},\ \bibinfo {pages} {8513} (\bibinfo {year}
  {2010})}\BibitemShut {NoStop}%
\bibitem [{\citenamefont {Riedel}\ \emph {et~al.}(2012)\citenamefont {Riedel},
  \citenamefont {Fuchs}, \citenamefont {Kraus}, \citenamefont {V\"ath},
  \citenamefont {Sperlich}, \citenamefont {Dyakonov}, \citenamefont
  {Soltamova}, \citenamefont {Baranov}, \citenamefont {Ilyin},\ and\
  \citenamefont {Astakhov}}]{Riedel2012}%
  \BibitemOpen
  \bibfield  {author} {\bibinfo {author} {\bibfnamefont {D.}~\bibnamefont
  {Riedel}}, \bibinfo {author} {\bibfnamefont {F.}~\bibnamefont {Fuchs}},
  \bibinfo {author} {\bibfnamefont {H.}~\bibnamefont {Kraus}}, \bibinfo
  {author} {\bibfnamefont {S.}~\bibnamefont {V\"ath}}, \bibinfo {author}
  {\bibfnamefont {A.}~\bibnamefont {Sperlich}}, \bibinfo {author}
  {\bibfnamefont {V.}~\bibnamefont {Dyakonov}}, \bibinfo {author}
  {\bibfnamefont {A.~A.}\ \bibnamefont {Soltamova}}, \bibinfo {author}
  {\bibfnamefont {P.~G.}\ \bibnamefont {Baranov}}, \bibinfo {author}
  {\bibfnamefont {V.~A.}\ \bibnamefont {Ilyin}}, \ and\ \bibinfo {author}
  {\bibfnamefont {G.~V.}\ \bibnamefont {Astakhov}},\ }\href {\doibase
  10.1103/PhysRevLett.109.226402} {\bibfield  {journal} {\bibinfo  {journal}
  {Phys. Rev. Lett.}\ }\textbf {\bibinfo {volume} {109}},\ \bibinfo {pages}
  {226402} (\bibinfo {year} {2012})}\BibitemShut {NoStop}%
\bibitem [{\citenamefont {Kraus}\ \emph {et~al.}(2014)\citenamefont {Kraus},
  \citenamefont {Soltamov}, \citenamefont {Riedel}, \citenamefont {V\"ath},
  \citenamefont {Fuchs}, \citenamefont {Sperlich}, \citenamefont {Baranov},
  \citenamefont {Dyakonov},\ and\ \citenamefont {Astakhov}}]{Kraus2014}%
  \BibitemOpen
  \bibfield  {author} {\bibinfo {author} {\bibfnamefont {H.}~\bibnamefont
  {Kraus}}, \bibinfo {author} {\bibfnamefont {V.~A.}\ \bibnamefont {Soltamov}},
  \bibinfo {author} {\bibfnamefont {D.}~\bibnamefont {Riedel}}, \bibinfo
  {author} {\bibfnamefont {S.}~\bibnamefont {V\"ath}}, \bibinfo {author}
  {\bibfnamefont {F.}~\bibnamefont {Fuchs}}, \bibinfo {author} {\bibfnamefont
  {A.}~\bibnamefont {Sperlich}}, \bibinfo {author} {\bibfnamefont {P.~G.}\
  \bibnamefont {Baranov}}, \bibinfo {author} {\bibfnamefont {V.}~\bibnamefont
  {Dyakonov}}, \ and\ \bibinfo {author} {\bibfnamefont {G.~V.}\ \bibnamefont
  {Astakhov}},\ }\href {\doibase 10.1038/nphys2826} {\bibfield  {journal}
  {\bibinfo  {journal} {Nat. Phys.}\ }\textbf {\bibinfo {volume} {10}},\
  \bibinfo {pages} {157} (\bibinfo {year} {2014})}\BibitemShut {NoStop}%
\bibitem [{\citenamefont {S\"orman}\ \emph {et~al.}(2000)\citenamefont
  {S\"orman}, \citenamefont {Son}, \citenamefont {Chen}, \citenamefont
  {Kordina}, \citenamefont {Hallin},\ and\ \citenamefont
  {Janz\'en}}]{Sorman2000}%
  \BibitemOpen
  \bibfield  {author} {\bibinfo {author} {\bibfnamefont {E.}~\bibnamefont
  {S\"orman}}, \bibinfo {author} {\bibfnamefont {N.~T.}\ \bibnamefont {Son}},
  \bibinfo {author} {\bibfnamefont {W.~M.}\ \bibnamefont {Chen}}, \bibinfo
  {author} {\bibfnamefont {O.}~\bibnamefont {Kordina}}, \bibinfo {author}
  {\bibfnamefont {C.}~\bibnamefont {Hallin}}, \ and\ \bibinfo {author}
  {\bibfnamefont {E.}~\bibnamefont {Janz\'en}},\ }\href {\doibase
  10.1103/PhysRevB.61.2613} {\bibfield  {journal} {\bibinfo  {journal} {Phys.
  Rev. B}\ }\textbf {\bibinfo {volume} {61}},\ \bibinfo {pages} {2613}
  (\bibinfo {year} {2000})}\BibitemShut {NoStop}%
\bibitem [{\citenamefont {Wagner}\ \emph {et~al.}(2000)\citenamefont {Wagner},
  \citenamefont {Magnusson}, \citenamefont {Chen}, \citenamefont {Janz\'en},
  \citenamefont {S\"orman}, \citenamefont {Hallin},\ and\ \citenamefont
  {Lindstr\"om}}]{Wagner2000}%
  \BibitemOpen
  \bibfield  {author} {\bibinfo {author} {\bibfnamefont {M.}~\bibnamefont
  {Wagner}}, \bibinfo {author} {\bibfnamefont {B.}~\bibnamefont {Magnusson}},
  \bibinfo {author} {\bibfnamefont {W.~M.}\ \bibnamefont {Chen}}, \bibinfo
  {author} {\bibfnamefont {E.}~\bibnamefont {Janz\'en}}, \bibinfo {author}
  {\bibfnamefont {E.}~\bibnamefont {S\"orman}}, \bibinfo {author}
  {\bibfnamefont {C.}~\bibnamefont {Hallin}}, \ and\ \bibinfo {author}
  {\bibfnamefont {J.~L.}\ \bibnamefont {Lindstr\"om}},\ }\href {\doibase
  10.1103/PhysRevB.62.16555} {\bibfield  {journal} {\bibinfo  {journal} {Phys.
  Rev. B}\ }\textbf {\bibinfo {volume} {62}},\ \bibinfo {pages} {16555}
  (\bibinfo {year} {2000})}\BibitemShut {NoStop}%
\bibitem [{\citenamefont {Lee}\ \emph {et~al.}(2015)\citenamefont {Lee},
  \citenamefont {Niethammer},\ and\ \citenamefont {Wrachtrup}}]{Lee2015}%
  \BibitemOpen
  \bibfield  {author} {\bibinfo {author} {\bibfnamefont {S.-Y.}\ \bibnamefont
  {Lee}}, \bibinfo {author} {\bibfnamefont {M.}~\bibnamefont {Niethammer}}, \
  and\ \bibinfo {author} {\bibfnamefont {J.}~\bibnamefont {Wrachtrup}},\ }\href
  {\doibase 10.1103/PhysRevB.92.115201} {\bibfield  {journal} {\bibinfo
  {journal} {Phys. Rev. B}\ }\textbf {\bibinfo {volume} {92}},\ \bibinfo
  {pages} {115201} (\bibinfo {year} {2015})}\BibitemShut {NoStop}%
\bibitem [{\citenamefont {Simin}\ \emph {et~al.}(2016)\citenamefont {Simin},
  \citenamefont {Soltamov}, \citenamefont {Poshakinskiy}, \citenamefont
  {Anisimov}, \citenamefont {Babunts}, \citenamefont {Tolmachev}, \citenamefont
  {Mokhov}, \citenamefont {Trupke}, \citenamefont {Tarasenko}, \citenamefont
  {Sperlich}, \citenamefont {Baranov}, \citenamefont {Dyakonov},\ and\
  \citenamefont {Astakhov}}]{Simin2016}%
  \BibitemOpen
  \bibfield  {author} {\bibinfo {author} {\bibfnamefont {D.}~\bibnamefont
  {Simin}}, \bibinfo {author} {\bibfnamefont {V.~A.}\ \bibnamefont {Soltamov}},
  \bibinfo {author} {\bibfnamefont {A.~V.}\ \bibnamefont {Poshakinskiy}},
  \bibinfo {author} {\bibfnamefont {A.~N.}\ \bibnamefont {Anisimov}}, \bibinfo
  {author} {\bibfnamefont {R.~A.}\ \bibnamefont {Babunts}}, \bibinfo {author}
  {\bibfnamefont {D.~O.}\ \bibnamefont {Tolmachev}}, \bibinfo {author}
  {\bibfnamefont {E.~N.}\ \bibnamefont {Mokhov}}, \bibinfo {author}
  {\bibfnamefont {M.}~\bibnamefont {Trupke}}, \bibinfo {author} {\bibfnamefont
  {S.~A.}\ \bibnamefont {Tarasenko}}, \bibinfo {author} {\bibfnamefont
  {A.}~\bibnamefont {Sperlich}}, \bibinfo {author} {\bibfnamefont {P.~G.}\
  \bibnamefont {Baranov}}, \bibinfo {author} {\bibfnamefont {V.}~\bibnamefont
  {Dyakonov}}, \ and\ \bibinfo {author} {\bibfnamefont {G.~V.}\ \bibnamefont
  {Astakhov}},\ }\href {\doibase 10.1103/PhysRevX.6.031014} {\bibfield
  {journal} {\bibinfo  {journal} {Phys. Rev. X}\ }\textbf {\bibinfo {volume}
  {6}},\ \bibinfo {pages} {031014} (\bibinfo {year} {2016})}\BibitemShut
  {NoStop}%
\bibitem [{\citenamefont {Niethammer}\ \emph {et~al.}(2016)\citenamefont
  {Niethammer}, \citenamefont {Widmann}, \citenamefont {Lee}, \citenamefont
  {Stenberg}, \citenamefont {Kordina}, \citenamefont {Ohshima}, \citenamefont
  {Son}, \citenamefont {Janz\'en},\ and\ \citenamefont
  {Wrachtrup}}]{Niethammer2016}%
  \BibitemOpen
  \bibfield  {author} {\bibinfo {author} {\bibfnamefont {M.}~\bibnamefont
  {Niethammer}}, \bibinfo {author} {\bibfnamefont {M.}~\bibnamefont {Widmann}},
  \bibinfo {author} {\bibfnamefont {S.-Y.}\ \bibnamefont {Lee}}, \bibinfo
  {author} {\bibfnamefont {P.}~\bibnamefont {Stenberg}}, \bibinfo {author}
  {\bibfnamefont {O.}~\bibnamefont {Kordina}}, \bibinfo {author} {\bibfnamefont
  {T.}~\bibnamefont {Ohshima}}, \bibinfo {author} {\bibfnamefont {N.~T.}\
  \bibnamefont {Son}}, \bibinfo {author} {\bibfnamefont {E.}~\bibnamefont
  {Janz\'en}}, \ and\ \bibinfo {author} {\bibfnamefont {J.}~\bibnamefont
  {Wrachtrup}},\ }\href {\doibase 10.1103/PhysRevApplied.6.034001} {\bibfield
  {journal} {\bibinfo  {journal} {Phys. Rev. Applied}\ }\textbf {\bibinfo
  {volume} {6}},\ \bibinfo {pages} {034001} (\bibinfo {year}
  {2016})}\BibitemShut {NoStop}%
\bibitem [{\citenamefont {Cochrane}\ \emph {et~al.}(2016)\citenamefont
  {Cochrane}, \citenamefont {Blacksberg}, \citenamefont {Anders},\ and\
  \citenamefont {Lenahan}}]{Cochrane2016}%
  \BibitemOpen
  \bibfield  {author} {\bibinfo {author} {\bibfnamefont {C.~J.}\ \bibnamefont
  {Cochrane}}, \bibinfo {author} {\bibfnamefont {J.}~\bibnamefont
  {Blacksberg}}, \bibinfo {author} {\bibfnamefont {M.~A.}\ \bibnamefont
  {Anders}}, \ and\ \bibinfo {author} {\bibfnamefont {P.~M.}\ \bibnamefont
  {Lenahan}},\ }\href {\doibase 10.1038/srep37077} {\bibfield  {journal}
  {\bibinfo  {journal} {Scientific Reports}\ }\textbf {\bibinfo {volume} {6}},\
  \bibinfo {pages} {srep37077} (\bibinfo {year} {2016})}\BibitemShut {NoStop}%
\bibitem [{\citenamefont {Anisimov}\ \emph {et~al.}(2016)\citenamefont
  {Anisimov}, \citenamefont {Simin}, \citenamefont {Soltamov}, \citenamefont
  {Lebedev}, \citenamefont {Baranov}, \citenamefont {Astakhov},\ and\
  \citenamefont {Dyakonov}}]{Anisimov2016}%
  \BibitemOpen
  \bibfield  {author} {\bibinfo {author} {\bibfnamefont {A.~N.}\ \bibnamefont
  {Anisimov}}, \bibinfo {author} {\bibfnamefont {D.}~\bibnamefont {Simin}},
  \bibinfo {author} {\bibfnamefont {V.~A.}\ \bibnamefont {Soltamov}}, \bibinfo
  {author} {\bibfnamefont {S.~P.}\ \bibnamefont {Lebedev}}, \bibinfo {author}
  {\bibfnamefont {P.~G.}\ \bibnamefont {Baranov}}, \bibinfo {author}
  {\bibfnamefont {G.~V.}\ \bibnamefont {Astakhov}}, \ and\ \bibinfo {author}
  {\bibfnamefont {V.}~\bibnamefont {Dyakonov}},\ }\href
  {http://dx.doi.org/10.1038/srep33301} {\bibfield  {journal} {\bibinfo
  {journal} {Sci. Rep.}\ }\textbf {\bibinfo {volume} {6}},\ \bibinfo {pages}
  {33301} (\bibinfo {year} {2016})}\BibitemShut {NoStop}%
\bibitem [{\citenamefont {Wimbauer}\ \emph {et~al.}(1997)\citenamefont
  {Wimbauer}, \citenamefont {Meyer}, \citenamefont {Hofstaetter}, \citenamefont
  {Scharmann},\ and\ \citenamefont {Overhof}}]{Wimbauer97}%
  \BibitemOpen
  \bibfield  {author} {\bibinfo {author} {\bibfnamefont {T.}~\bibnamefont
  {Wimbauer}}, \bibinfo {author} {\bibfnamefont {B.~K.}\ \bibnamefont {Meyer}},
  \bibinfo {author} {\bibfnamefont {A.}~\bibnamefont {Hofstaetter}}, \bibinfo
  {author} {\bibfnamefont {A.}~\bibnamefont {Scharmann}}, \ and\ \bibinfo
  {author} {\bibfnamefont {H.}~\bibnamefont {Overhof}},\ }\href {\doibase
  10.1103/PhysRevB.56.7384} {\bibfield  {journal} {\bibinfo  {journal} {Phys.
  Rev. B}\ }\textbf {\bibinfo {volume} {56}},\ \bibinfo {pages} {7384}
  (\bibinfo {year} {1997})}\BibitemShut {NoStop}%
\bibitem [{\citenamefont {Mizuochi}\ \emph {et~al.}(2002)\citenamefont
  {Mizuochi}, \citenamefont {Yamasaki}, \citenamefont {Takizawa}, \citenamefont
  {Morishita}, \citenamefont {Ohshima}, \citenamefont {Itoh},\ and\
  \citenamefont {Isoya}}]{Mizuochi2002}%
  \BibitemOpen
  \bibfield  {author} {\bibinfo {author} {\bibfnamefont {N.}~\bibnamefont
  {Mizuochi}}, \bibinfo {author} {\bibfnamefont {S.}~\bibnamefont {Yamasaki}},
  \bibinfo {author} {\bibfnamefont {H.}~\bibnamefont {Takizawa}}, \bibinfo
  {author} {\bibfnamefont {N.}~\bibnamefont {Morishita}}, \bibinfo {author}
  {\bibfnamefont {T.}~\bibnamefont {Ohshima}}, \bibinfo {author} {\bibfnamefont
  {H.}~\bibnamefont {Itoh}}, \ and\ \bibinfo {author} {\bibfnamefont
  {J.}~\bibnamefont {Isoya}},\ }\href {\doibase 10.1103/PhysRevB.66.235202}
  {\bibfield  {journal} {\bibinfo  {journal} {Phys. Rev. B}\ }\textbf {\bibinfo
  {volume} {66}},\ \bibinfo {pages} {235202} (\bibinfo {year}
  {2002})}\BibitemShut {NoStop}%
\bibitem [{\citenamefont {Orlinski}\ \emph {et~al.}(2003)\citenamefont
  {Orlinski}, \citenamefont {Schmidt}, \citenamefont {Mokhov},\ and\
  \citenamefont {Baranov}}]{Orlinski2003}%
  \BibitemOpen
  \bibfield  {author} {\bibinfo {author} {\bibfnamefont {S.~B.}\ \bibnamefont
  {Orlinski}}, \bibinfo {author} {\bibfnamefont {J.}~\bibnamefont {Schmidt}},
  \bibinfo {author} {\bibfnamefont {E.~N.}\ \bibnamefont {Mokhov}}, \ and\
  \bibinfo {author} {\bibfnamefont {P.~G.}\ \bibnamefont {Baranov}},\ }\href
  {\doibase 10.1103/PhysRevB.67.125207} {\bibfield  {journal} {\bibinfo
  {journal} {Phys. Rev. B}\ }\textbf {\bibinfo {volume} {67}},\ \bibinfo
  {pages} {125207} (\bibinfo {year} {2003})}\BibitemShut {NoStop}%
\bibitem [{\citenamefont {Mizuochi}\ \emph {et~al.}(2005)\citenamefont
  {Mizuochi}, \citenamefont {Yamasaki}, \citenamefont {Takizawa}, \citenamefont
  {Morishita}, \citenamefont {Ohshima}, \citenamefont {Itoh}, \citenamefont
  {Umeda},\ and\ \citenamefont {Isoya}}]{Mizuochi2005}%
  \BibitemOpen
  \bibfield  {author} {\bibinfo {author} {\bibfnamefont {N.}~\bibnamefont
  {Mizuochi}}, \bibinfo {author} {\bibfnamefont {S.}~\bibnamefont {Yamasaki}},
  \bibinfo {author} {\bibfnamefont {H.}~\bibnamefont {Takizawa}}, \bibinfo
  {author} {\bibfnamefont {N.}~\bibnamefont {Morishita}}, \bibinfo {author}
  {\bibfnamefont {T.}~\bibnamefont {Ohshima}}, \bibinfo {author} {\bibfnamefont
  {H.}~\bibnamefont {Itoh}}, \bibinfo {author} {\bibfnamefont {T.}~\bibnamefont
  {Umeda}}, \ and\ \bibinfo {author} {\bibfnamefont {J.}~\bibnamefont
  {Isoya}},\ }\href {\doibase 10.1103/PhysRevB.72.235208} {\bibfield  {journal}
  {\bibinfo  {journal} {Phys. Rev. B}\ }\textbf {\bibinfo {volume} {72}},\
  \bibinfo {pages} {235208} (\bibinfo {year} {2005})}\BibitemShut {NoStop}%
\bibitem [{\citenamefont {Janz\'en}\ \emph {et~al.}(2009)\citenamefont
  {Janz\'en}, \citenamefont {Gali}, \citenamefont {Carlsson}, \citenamefont
  {G\"allstr\"om}, \citenamefont {Magnusson},\ and\ \citenamefont
  {Son}}]{Janzen2009}%
  \BibitemOpen
  \bibfield  {author} {\bibinfo {author} {\bibfnamefont {E.}~\bibnamefont
  {Janz\'en}}, \bibinfo {author} {\bibfnamefont {A.}~\bibnamefont {Gali}},
  \bibinfo {author} {\bibfnamefont {P.}~\bibnamefont {Carlsson}}, \bibinfo
  {author} {\bibfnamefont {A.}~\bibnamefont {G\"allstr\"om}}, \bibinfo {author}
  {\bibfnamefont {B.}~\bibnamefont {Magnusson}}, \ and\ \bibinfo {author}
  {\bibfnamefont {N.}~\bibnamefont {Son}},\ }\href {\doibase
  https://doi.org/10.1016/j.physb.2009.09.023} {\bibfield  {journal} {\bibinfo
  {journal} {Physica B: Condensed Matter}\ }\textbf {\bibinfo {volume} {404}},\
  \bibinfo {pages} {4354 } (\bibinfo {year} {2009})},\ \bibinfo {note}
  {proceedings of the Third South African Conference on Photonic
  Materials}\BibitemShut {NoStop}%
\bibitem [{\citenamefont {Soltamov}\ \emph {et~al.}(2015)\citenamefont
  {Soltamov}, \citenamefont {Yavkin}, \citenamefont {Tolmachev}, \citenamefont
  {Babunts}, \citenamefont {Badalyan}, \citenamefont {Davydov}, \citenamefont
  {Mokhov}, \citenamefont {Proskuryakov}, \citenamefont {Orlinskii},\ and\
  \citenamefont {Baranov}}]{Soltamov2015}%
  \BibitemOpen
  \bibfield  {author} {\bibinfo {author} {\bibfnamefont {V.~A.}\ \bibnamefont
  {Soltamov}}, \bibinfo {author} {\bibfnamefont {B.~V.}\ \bibnamefont
  {Yavkin}}, \bibinfo {author} {\bibfnamefont {D.~O.}\ \bibnamefont
  {Tolmachev}}, \bibinfo {author} {\bibfnamefont {R.~A.}\ \bibnamefont
  {Babunts}}, \bibinfo {author} {\bibfnamefont {A.~G.}\ \bibnamefont
  {Badalyan}}, \bibinfo {author} {\bibfnamefont {V.~Y.}\ \bibnamefont
  {Davydov}}, \bibinfo {author} {\bibfnamefont {E.~N.}\ \bibnamefont {Mokhov}},
  \bibinfo {author} {\bibfnamefont {I.~I.}\ \bibnamefont {Proskuryakov}},
  \bibinfo {author} {\bibfnamefont {S.~B.}\ \bibnamefont {Orlinskii}}, \ and\
  \bibinfo {author} {\bibfnamefont {P.~G.}\ \bibnamefont {Baranov}},\ }\href
  {\doibase 10.1103/PhysRevLett.115.247602} {\bibfield  {journal} {\bibinfo
  {journal} {Phys. Rev. Lett.}\ }\textbf {\bibinfo {volume} {115}},\ \bibinfo
  {pages} {247602} (\bibinfo {year} {2015})}\BibitemShut {NoStop}%
\bibitem [{\citenamefont {Soykal}\ \emph {et~al.}(2016)\citenamefont {Soykal},
  \citenamefont {Dev},\ and\ \citenamefont {Economou}}]{Soykal2016}%
  \BibitemOpen
  \bibfield  {author} {\bibinfo {author} {\bibfnamefont {O.~O.}\ \bibnamefont
  {Soykal}}, \bibinfo {author} {\bibfnamefont {P.}~\bibnamefont {Dev}}, \ and\
  \bibinfo {author} {\bibfnamefont {S.~E.}\ \bibnamefont {Economou}},\ }\href
  {\doibase 10.1103/PhysRevB.93.081207} {\bibfield  {journal} {\bibinfo
  {journal} {Phys. Rev. B}\ }\textbf {\bibinfo {volume} {93}},\ \bibinfo
  {pages} {081207} (\bibinfo {year} {2016})}\BibitemShut {NoStop}%
\bibitem [{\citenamefont {Mizuochi}\ \emph {et~al.}(2003)\citenamefont
  {Mizuochi}, \citenamefont {Yamasaki}, \citenamefont {Takizawa}, \citenamefont
  {Morishita}, \citenamefont {Ohshima}, \citenamefont {Itoh},\ and\
  \citenamefont {Isoya}}]{Mizuochi2003}%
  \BibitemOpen
  \bibfield  {author} {\bibinfo {author} {\bibfnamefont {N.}~\bibnamefont
  {Mizuochi}}, \bibinfo {author} {\bibfnamefont {S.}~\bibnamefont {Yamasaki}},
  \bibinfo {author} {\bibfnamefont {H.}~\bibnamefont {Takizawa}}, \bibinfo
  {author} {\bibfnamefont {N.}~\bibnamefont {Morishita}}, \bibinfo {author}
  {\bibfnamefont {T.}~\bibnamefont {Ohshima}}, \bibinfo {author} {\bibfnamefont
  {H.}~\bibnamefont {Itoh}}, \ and\ \bibinfo {author} {\bibfnamefont
  {J.}~\bibnamefont {Isoya}},\ }\href {\doibase 10.1103/PhysRevB.68.165206}
  {\bibfield  {journal} {\bibinfo  {journal} {Phys. Rev. B}\ }\textbf {\bibinfo
  {volume} {68}},\ \bibinfo {pages} {165206} (\bibinfo {year}
  {2003})}\BibitemShut {NoStop}%
\bibitem [{\citenamefont {Bl\"ochl}(1994)}]{PAW}%
  \BibitemOpen
  \bibfield  {author} {\bibinfo {author} {\bibfnamefont {P.~E.}\ \bibnamefont
  {Bl\"ochl}},\ }\href {\doibase 10.1103/PhysRevB.50.17953} {\bibfield
  {journal} {\bibinfo  {journal} {Phys. Rev. B}\ }\textbf {\bibinfo {volume}
  {50}},\ \bibinfo {pages} {17953} (\bibinfo {year} {1994})}\BibitemShut
  {NoStop}%
\bibitem [{\citenamefont {Kresse}\ and\ \citenamefont {Hafner}(1994)}]{VASP}%
  \BibitemOpen
  \bibfield  {author} {\bibinfo {author} {\bibfnamefont {G.}~\bibnamefont
  {Kresse}}\ and\ \bibinfo {author} {\bibfnamefont {J.}~\bibnamefont
  {Hafner}},\ }\href {\doibase 10.1103/PhysRevB.49.14251} {\bibfield  {journal}
  {\bibinfo  {journal} {Phys. Rev. B}\ }\textbf {\bibinfo {volume} {49}},\
  \bibinfo {pages} {14251} (\bibinfo {year} {1994})}\BibitemShut {NoStop}%
\bibitem [{\citenamefont {Kresse}\ and\ \citenamefont
  {Furthm\"uller}(1996)}]{VASP2}%
  \BibitemOpen
  \bibfield  {author} {\bibinfo {author} {\bibfnamefont {G.}~\bibnamefont
  {Kresse}}\ and\ \bibinfo {author} {\bibfnamefont {J.}~\bibnamefont
  {Furthm\"uller}},\ }\href {\doibase 10.1103/PhysRevB.54.11169} {\bibfield
  {journal} {\bibinfo  {journal} {Phys. Rev. B}\ }\textbf {\bibinfo {volume}
  {54}},\ \bibinfo {pages} {11169} (\bibinfo {year} {1996})}\BibitemShut
  {NoStop}%
\bibitem [{\citenamefont {Heyd}\ \emph {et~al.}(2003)\citenamefont {Heyd},
  \citenamefont {Scuseria},\ and\ \citenamefont {Ernzerhof}}]{HSE03}%
  \BibitemOpen
  \bibfield  {author} {\bibinfo {author} {\bibfnamefont {J.}~\bibnamefont
  {Heyd}}, \bibinfo {author} {\bibfnamefont {G.~E.}\ \bibnamefont {Scuseria}},
  \ and\ \bibinfo {author} {\bibfnamefont {M.}~\bibnamefont {Ernzerhof}},\
  }\href {\doibase 10.1063/1.1564060} {\bibfield  {journal} {\bibinfo
  {journal} {J. Chem. Phys.}\ }\textbf {\bibinfo {volume} {118}},\ \bibinfo
  {pages} {8207} (\bibinfo {year} {2003})}\BibitemShut {NoStop}%
\bibitem [{\citenamefont {Heyd}\ \emph {et~al.}(2006)\citenamefont {Heyd},
  \citenamefont {Scuseria},\ and\ \citenamefont {Ernzerhof}}]{HSE06}%
  \BibitemOpen
  \bibfield  {author} {\bibinfo {author} {\bibfnamefont {J.}~\bibnamefont
  {Heyd}}, \bibinfo {author} {\bibfnamefont {G.~E.}\ \bibnamefont {Scuseria}},
  \ and\ \bibinfo {author} {\bibfnamefont {M.}~\bibnamefont {Ernzerhof}},\
  }\href {\doibase 10.1063/1.2204597} {\bibfield  {journal} {\bibinfo
  {journal} {J. Chem. Phys.}\ }\textbf {\bibinfo {volume} {124}},\ \bibinfo
  {pages} {219906} (\bibinfo {year} {2006})}\BibitemShut {NoStop}%
\bibitem [{\citenamefont {Gali}(2009)}]{Gali2009}%
  \BibitemOpen
  \bibfield  {author} {\bibinfo {author} {\bibfnamefont {A.}~\bibnamefont
  {Gali}},\ }\href {\doibase 10.1103/PhysRevB.80.241204} {\bibfield  {journal}
  {\bibinfo  {journal} {Phys. Rev. B}\ }\textbf {\bibinfo {volume} {80}},\
  \bibinfo {pages} {241204} (\bibinfo {year} {2009})}\BibitemShut {NoStop}%
\bibitem [{\citenamefont {De\'ak}\ \emph {et~al.}(2010)\citenamefont {De\'ak},
  \citenamefont {Aradi}, \citenamefont {Frauenheim}, \citenamefont {Janz\'en},\
  and\ \citenamefont {Gali}}]{Deak2010}%
  \BibitemOpen
  \bibfield  {author} {\bibinfo {author} {\bibfnamefont {P.}~\bibnamefont
  {De\'ak}}, \bibinfo {author} {\bibfnamefont {B.}~\bibnamefont {Aradi}},
  \bibinfo {author} {\bibfnamefont {T.}~\bibnamefont {Frauenheim}}, \bibinfo
  {author} {\bibfnamefont {E.}~\bibnamefont {Janz\'en}}, \ and\ \bibinfo
  {author} {\bibfnamefont {A.}~\bibnamefont {Gali}},\ }\href {\doibase
  10.1103/PhysRevB.81.153203} {\bibfield  {journal} {\bibinfo  {journal} {Phys.
  Rev. B}\ }\textbf {\bibinfo {volume} {81}},\ \bibinfo {pages} {153203}
  (\bibinfo {year} {2010})}\BibitemShut {NoStop}%
\bibitem [{\citenamefont {Sz\'asz}\ \emph {et~al.}(2013)\citenamefont
  {Sz\'asz}, \citenamefont {Hornos}, \citenamefont {Marsman},\ and\
  \citenamefont {Gali}}]{Szasz2013}%
  \BibitemOpen
  \bibfield  {author} {\bibinfo {author} {\bibfnamefont {K.}~\bibnamefont
  {Sz\'asz}}, \bibinfo {author} {\bibfnamefont {T.}~\bibnamefont {Hornos}},
  \bibinfo {author} {\bibfnamefont {M.}~\bibnamefont {Marsman}}, \ and\
  \bibinfo {author} {\bibfnamefont {A.}~\bibnamefont {Gali}},\ }\href {\doibase
  10.1103/PhysRevB.88.075202} {\bibfield  {journal} {\bibinfo  {journal} {Phys.
  Rev. B}\ }\textbf {\bibinfo {volume} {88}},\ \bibinfo {pages} {075202}
  (\bibinfo {year} {2013})}\BibitemShut {NoStop}%
\bibitem [{\citenamefont {Perdew}\ \emph {et~al.}(1996)\citenamefont {Perdew},
  \citenamefont {Burke},\ and\ \citenamefont {Ernzerhof}}]{PBE}%
  \BibitemOpen
  \bibfield  {author} {\bibinfo {author} {\bibfnamefont {J.~P.}\ \bibnamefont
  {Perdew}}, \bibinfo {author} {\bibfnamefont {K.}~\bibnamefont {Burke}}, \
  and\ \bibinfo {author} {\bibfnamefont {M.}~\bibnamefont {Ernzerhof}},\ }\href
  {\doibase 10.1103/PhysRevLett.77.3865} {\bibfield  {journal} {\bibinfo
  {journal} {Phys. Rev. Lett.}\ }\textbf {\bibinfo {volume} {77}},\ \bibinfo
  {pages} {3865} (\bibinfo {year} {1996})}\BibitemShut {NoStop}%
\bibitem [{\citenamefont {Iv\'ady}\ \emph {et~al.}(2014)\citenamefont
  {Iv\'ady}, \citenamefont {Simon}, \citenamefont {Maze}, \citenamefont
  {Abrikosov},\ and\ \citenamefont {Gali}}]{Ivady2014}%
  \BibitemOpen
  \bibfield  {author} {\bibinfo {author} {\bibfnamefont {V.}~\bibnamefont
  {Iv\'ady}}, \bibinfo {author} {\bibfnamefont {T.}~\bibnamefont {Simon}},
  \bibinfo {author} {\bibfnamefont {J.~R.}\ \bibnamefont {Maze}}, \bibinfo
  {author} {\bibfnamefont {I.~A.}\ \bibnamefont {Abrikosov}}, \ and\ \bibinfo
  {author} {\bibfnamefont {A.}~\bibnamefont {Gali}},\ }\href@noop {} {\bibfield
   {journal} {\bibinfo  {journal} {Phys. Rev. B}\ }\textbf {\bibinfo {volume}
  {90}},\ \bibinfo {pages} {235205} (\bibinfo {year} {2014})}\BibitemShut
  {NoStop}%
\bibitem [{\citenamefont {Falk}\ \emph {et~al.}(2014)\citenamefont {Falk},
  \citenamefont {Klimov}, \citenamefont {Buckley}, \citenamefont {Iv\'ady},
  \citenamefont {Abrikosov}, \citenamefont {Calusine}, \citenamefont {Koehl},
  \citenamefont {Gali},\ and\ \citenamefont {Awschalom}}]{Falk2014}%
  \BibitemOpen
  \bibfield  {author} {\bibinfo {author} {\bibfnamefont {A.~L.}\ \bibnamefont
  {Falk}}, \bibinfo {author} {\bibfnamefont {P.~V.}\ \bibnamefont {Klimov}},
  \bibinfo {author} {\bibfnamefont {B.~B.}\ \bibnamefont {Buckley}}, \bibinfo
  {author} {\bibfnamefont {V.}~\bibnamefont {Iv\'ady}}, \bibinfo {author}
  {\bibfnamefont {I.~A.}\ \bibnamefont {Abrikosov}}, \bibinfo {author}
  {\bibfnamefont {G.}~\bibnamefont {Calusine}}, \bibinfo {author}
  {\bibfnamefont {W.~F.}\ \bibnamefont {Koehl}}, \bibinfo {author}
  {\bibfnamefont {A.}~\bibnamefont {Gali}}, \ and\ \bibinfo {author}
  {\bibfnamefont {D.~D.}\ \bibnamefont {Awschalom}},\ }\href {\doibase
  10.1103/PhysRevLett.112.187601} {\bibfield  {journal} {\bibinfo  {journal}
  {Phys. Rev. Lett.}\ }\textbf {\bibinfo {volume} {112}},\ \bibinfo {pages}
  {187601} (\bibinfo {year} {2014})}\BibitemShut {NoStop}%
\bibitem [{\citenamefont {Davidsson}\ \emph {et~al.}(2017)\citenamefont
  {Davidsson}, \citenamefont {Iv\'ady}, \citenamefont {Armiento}, \citenamefont
  {Son}, \citenamefont {Gali},\ and\ \citenamefont {Abrikosov}}]{Joel2017}%
  \BibitemOpen
  \bibfield  {author} {\bibinfo {author} {\bibfnamefont {J.}~\bibnamefont
  {Davidsson}}, \bibinfo {author} {\bibfnamefont {V.}~\bibnamefont {Iv\'ady}},
  \bibinfo {author} {\bibfnamefont {R.}~\bibnamefont {Armiento}}, \bibinfo
  {author} {\bibfnamefont {N.~T.}\ \bibnamefont {Son}}, \bibinfo {author}
  {\bibfnamefont {A.}~\bibnamefont {Gali}}, \ and\ \bibinfo {author}
  {\bibfnamefont {I.~A.}\ \bibnamefont {Abrikosov}},\ }\href@noop {} {\bibfield
   {journal} {\bibinfo  {journal} {arXiv:1708.04508}\ } (\bibinfo {year}
  {2017})}\BibitemShut {NoStop}%
\bibitem [{Not({\natexlab{a}})}]{Note1}%
  \BibitemOpen
  \href@noop {} {}\bibinfo {howpublished} {See Supplementary Materials for
  information about the conditions of EPR measurements and additional data in
  $6H$-SiC [url]} ({\natexlab{a}})\BibitemShut {NoStop}%
\bibitem [{\citenamefont {Trinh}\ \emph {et~al.}(2013)\citenamefont {Trinh},
  \citenamefont {Sz\'asz}, \citenamefont {Hornos}, \citenamefont {Kawahara},
  \citenamefont {Suda}, \citenamefont {Kimoto}, \citenamefont {Gali},
  \citenamefont {Janz\'en},\ and\ \citenamefont {Son}}]{Trinh2013}%
  \BibitemOpen
  \bibfield  {author} {\bibinfo {author} {\bibfnamefont {X.~T.}\ \bibnamefont
  {Trinh}}, \bibinfo {author} {\bibfnamefont {K.}~\bibnamefont {Sz\'asz}},
  \bibinfo {author} {\bibfnamefont {T.}~\bibnamefont {Hornos}}, \bibinfo
  {author} {\bibfnamefont {K.}~\bibnamefont {Kawahara}}, \bibinfo {author}
  {\bibfnamefont {J.}~\bibnamefont {Suda}}, \bibinfo {author} {\bibfnamefont
  {T.}~\bibnamefont {Kimoto}}, \bibinfo {author} {\bibfnamefont
  {A.}~\bibnamefont {Gali}}, \bibinfo {author} {\bibfnamefont {E.}~\bibnamefont
  {Janz\'en}}, \ and\ \bibinfo {author} {\bibfnamefont {N.~T.}\ \bibnamefont
  {Son}},\ }\href {\doibase 10.1103/PhysRevB.88.235209} {\bibfield  {journal}
  {\bibinfo  {journal} {Phys. Rev. B}\ }\textbf {\bibinfo {volume} {88}},\
  \bibinfo {pages} {235209} (\bibinfo {year} {2013})}\BibitemShut {NoStop}%
\bibitem [{\citenamefont {Son}\ \emph {et~al.}(2007)\citenamefont {Son},
  \citenamefont {Carlsson}, \citenamefont {ul~Hassan}, \citenamefont
  {Magnusson},\ and\ \citenamefont {Janz\'en}}]{Son2007}%
  \BibitemOpen
  \bibfield  {author} {\bibinfo {author} {\bibfnamefont {N.~T.}\ \bibnamefont
  {Son}}, \bibinfo {author} {\bibfnamefont {P.}~\bibnamefont {Carlsson}},
  \bibinfo {author} {\bibfnamefont {J.}~\bibnamefont {ul~Hassan}}, \bibinfo
  {author} {\bibfnamefont {B.}~\bibnamefont {Magnusson}}, \ and\ \bibinfo
  {author} {\bibfnamefont {E.}~\bibnamefont {Janz\'en}},\ }\href {\doibase
  10.1103/PhysRevB.75.155204} {\bibfield  {journal} {\bibinfo  {journal} {Phys.
  Rev. B}\ }\textbf {\bibinfo {volume} {75}},\ \bibinfo {pages} {155204}
  (\bibinfo {year} {2007})}\BibitemShut {NoStop}%
\bibitem [{Not({\natexlab{b}})}]{Note2}%
  \BibitemOpen
  \href@noop {} {}\bibinfo {howpublished} {The hyperfine constants of
  $\text{C}_{\text{Ia}}$ nuclei are within 0.6\% associated with the two sites,
  which is too small difference for decisive conclusion. However, this
  difference is significant for the other nuclei.} ({\natexlab{b}})\BibitemShut
  {NoStop}%
\bibitem [{\citenamefont {Sz\'asz}\ \emph {et~al.}(2014)\citenamefont
  {Sz\'asz}, \citenamefont {Trinh}, \citenamefont {Son}, \citenamefont
  {Janz\'en},\ and\ \citenamefont {Gali}}]{Szasz2014}%
  \BibitemOpen
  \bibfield  {author} {\bibinfo {author} {\bibfnamefont {K.}~\bibnamefont
  {Sz\'asz}}, \bibinfo {author} {\bibfnamefont {X.~T.}\ \bibnamefont {Trinh}},
  \bibinfo {author} {\bibfnamefont {N.~T.}\ \bibnamefont {Son}}, \bibinfo
  {author} {\bibfnamefont {E.}~\bibnamefont {Janz\'en}}, \ and\ \bibinfo
  {author} {\bibfnamefont {A.}~\bibnamefont {Gali}},\ }\href {\doibase
  http://dx.doi.org/10.1063/1.4866331} {\bibfield  {journal} {\bibinfo
  {journal} {J. Appl. Phys.}\ }\textbf {\bibinfo {volume} {115}},\ \bibinfo
  {eid} {073705} (\bibinfo {year} {2014}),\
  http://dx.doi.org/10.1063/1.4866331}\BibitemShut {NoStop}%
\end{thebibliography}

%

\end{document}